\documentclass[12pt,a4paper]{scrartcl}
\usepackage[left=2cm, right=2cm, top=2cm, bottom=30mm]{geometry}
\usepackage{etoolbox}
\usepackage{underscore,xspace,graphicx}
\usepackage{subfigure,caption}
\usepackage{amsmath,amssymb}
\usepackage{tikz,url}
\usepackage[squaren]{SIunits}
\usepackage{booktabs}
\usepackage{relsize}

\usepackage{titling}
\pretitle{\begin{flushleft}\LARGE\bfseries\sffamily}
\posttitle{\end{flushleft}\vspace*{-2.5mm}}
\predate{\begin{flushleft}\sffamily}
\postdate{\end{flushleft}}
\preauthor{\begin{flushleft}\large\sffamily}
\postauthor{\end{flushleft}}

\newcommand{\taxon}[1]{\textit{#1}}

\author{Todd C. Rae$^{\,1}$ and Andy Buckley$^{\,2}$}
\title{TaxMan: an online facility for the coding of continuous characters for cladistic analysis}
\date{12 July 2009, unpublished; \emph{uploaded to arXiv Nov 2014}}

\begin{document}
\maketitle
\thispagestyle{empty}

\noindent
{\smaller $^1$
  Centre for Research in Evolutionary Anthropology,\\
  School of Human and Life Sciences, Roehampton University,\\
  Holybourne Avenue, London SW15~4JD,\\
  UNITED KINGDOM;\\
  t.rae@roehampton.ac.uk}\\[1em]
{\smaller $^2$
  School of Physics \& Astronomy,\\
  University of Glasgow, Glasgow G12~7QQ,\\
  UNITED KINGDOM;\\
  \textit{formerly at}\\
  Centre for Particle Theory, Department of Physics,\\
  Durham University, South Road, Durham DH1~3LE,\\
  UNITED KINGDOM;\\

}

\vspace*{8mm}


\begin{minipage}{0.9\textwidth}
  \setlength{\parskip}{1ex}

  \quad
  A consensus is emerging that continuous (or metric) measures can be useful in
  phylogenetic systematics.  Many of the methods for coding such characters,
  however, employ elements that are arbitrary and therefore should be excluded
  from use in cladistic analysis.  The continued use of such potentially
  inappropriate methods can be attributed to either their simplicity or the
  availability of computer programs specifically designed to produce data
  matrices using these methods.  Conversely, one of the most suitable methods,
  homogeneous subset coding (HSC), is underused, probably due to the lack of a
  suitable software implementation for this somewhat complex procedure.

  This paper describes TaxMan, a Web-based facility for the coding of continuous
  data using HSC.  Data are entered using a form accessible via any internet
  browser and are automatically converted to a matrix suitable for input into
  tree-searching software.  This implementation of the HSC technique provides an
  uncomplicated procedure for the incorporation of metric data in phylogenetic
  systematics. The algorithmic implementation of the HSC procedure, and
  interpolation of the Studentised range and maximum modulus distributions
  required by it, is described in detail in appendices.
\end{minipage}

\clearpage

\tableofcontents

\section{Introduction}

\begin{center}
  \vspace*{7mm}
  \begin{minipage}[c]{0.7\linewidth}
    Human nature is, above all things, lazy.\\
    \hspace*{1cm}\textit{--- Harriet Beecher Stowe,}\\
    \hspace*{1cm}\textit{\hphantom{--- }Household Papers and Stories, 1864, ch. 6}
  \end{minipage}
  \vspace*{7mm}
\end{center}

It is generally agreed that the bulk of morphological variation in organisms is
quantitative, even if it is described at times as if it were meristic or
qualitative (Pogue~and~Mickevich,~1990; Stevens,~1991).  Although objections to
the use of quantitative traits in phylogenetic systematics have been raised in
the past (Crisp~and~Weston,~1987; Pimentel~and~Riggins,~1987), these have been
shown to be spurious (Rae,~1998).  There is an emerging consensus that
quantitative variation can (and should) be used in cladistics
(Garcia-Cruz~and~Sosa,~2006), but there is no agreement on the method(s) that
should be used to translate numerical data into data matrices amenable to
computer-based analysis, and several different methods have been suggested
(Simon,~1983; Thiele,~1993; Strait~et~al.,~1996; Goloboff~et~al.,~2004).

To a greater or lesser extent, however, the bulk of the methods proposed to date
for coding quantitative data for cladistic analysis include arbitrary aspects
(Rae,~1998), which limits their applicability to scientific enquiry.  This
objection applies to all forms of segment- or gap-coding, where there is no
objective \textit{a priori} criterion for deciding segment size or gap size
threshold.  All previously published ``tests'' of the efficacy of different
methods of coding rely on comparisons between topologies produced by various
coding methods (or continuous vs. ``non-continuous'' data); that the choice of
coding method can affect the resulting cladograms is a given, but comparing
trees does not reveal whether any particular coding method accurately reflects
variation.  For example, while results of comparisons between identical data
coded using different systems may reveal that one method produces a ``small
number of well-resolved trees with strong bootstrap support''
(Garcia-Cruz~and~Sosa,~2006:~302), this does necessarily mean that the results
reflect the evolution of the organisms accurately; the resultant ``good''
behaviour may be simply an incidental by-product of the method.  For example, it
may be that a particular method simply creates more character states (and hence
more apparent resolution) because the number of states created is arbitrarily
equal to the maximum number of states allowed by the tree-searching software
used (Thiele,~1993); thus, there could be more distinctions between taxa, and
perhaps even fewer equally parsimonious minimum-length trees, but the
distinctions would be illusory.  In addition, it has been shown
(Mickevich~and~Farris,~1981) that methods which add an element of scaling to the
coding procedure treat every character as if it has changed by the same
(arbitrary) degree as every other trait, a situation not likely to have occurred
many times in the course of the evolution of life.

Coding methods based on statistical tests between taxa suffer from no such
drawbacks.  There is no arbitrary (and thus infinitely variable) gap size
threshold and the number of character states is determined by the statistical
differences between taxa, not by the limitations of phylogenetic software
(Rae,~1998).  Surprisingly, however, these techniques are rarely used, probably
because they are difficult to implement using existing computer programs.

On the other hand, gap-weighting (Thiele~and~Ladiges,~1988) is a popular method
for coding quantitative characters (e.g., Torres-Carvajal,~2007), despite the
objections against it (see above).  This may be due to the fact that there is
software (Morphocode) that is specifically designed to translate metric data
into Nexus-format files (Schols~et~al.,~2004).  A recent implementation of a
modified version of gap-coding method is now available in TNT, as well
(Goloboff~et~al.,~2004); this method, although touted as treating ``continuous
characters as such'' (Goloboff~et~al.,~2004:~589), simply assigns a step (or
part thereof) corresponding to the distance between means or where the
``ranges'' of taxa do not overlap.  In this case, the arbitrary element is not
introduced by gap size, but by the characterisation of the range, which is ``up
to the user'' (Goloboff~et~al.,~2004:~591).  If the advent of computer
algorithms has taught us nothing else, it is that the available software will be
applied to a problem regardless of whether it is appropriate; hence the
continued reporting of majority-rule consensus trees (e.g.,
Belfiore~et~al.,~2008) long after they have been shown to be suspect from first
principles (Sharkey~and~Leathers,~2001).

\begin{figure}
  \centering
  \includegraphics[width=0.45\textwidth]{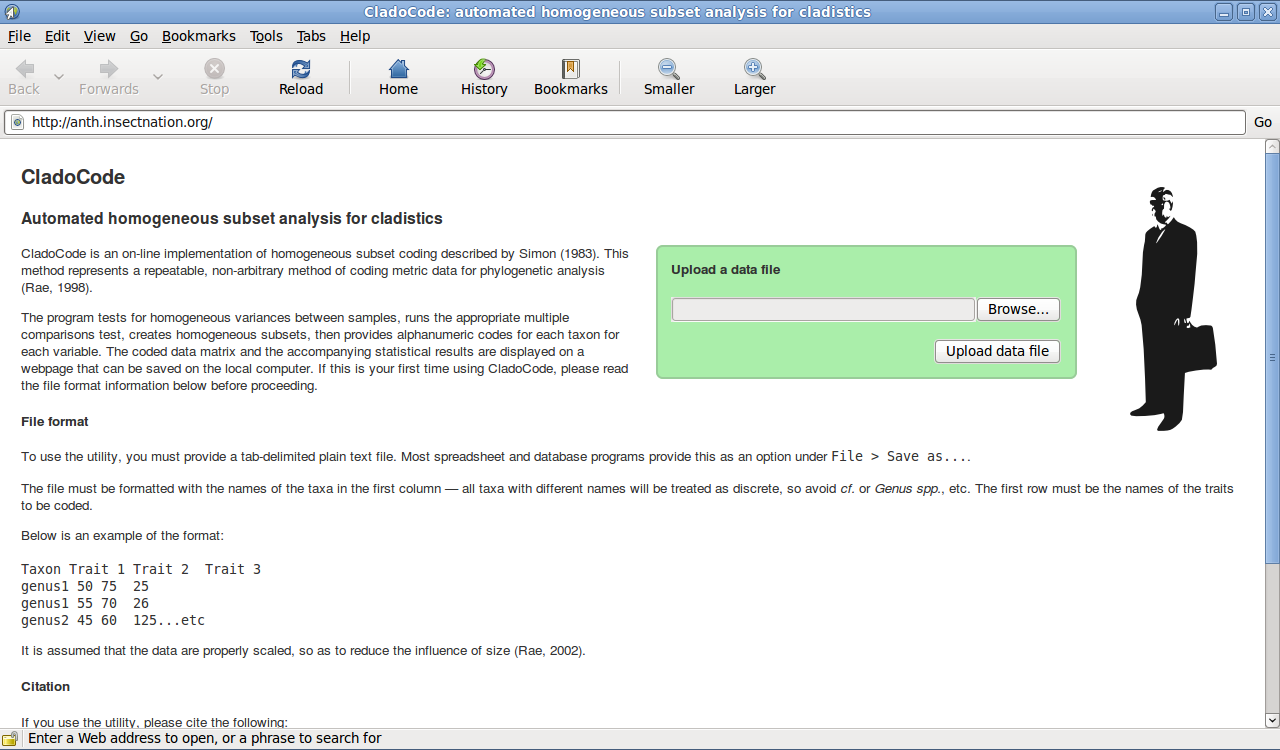}\qquad
  \includegraphics[width=0.45\textwidth]{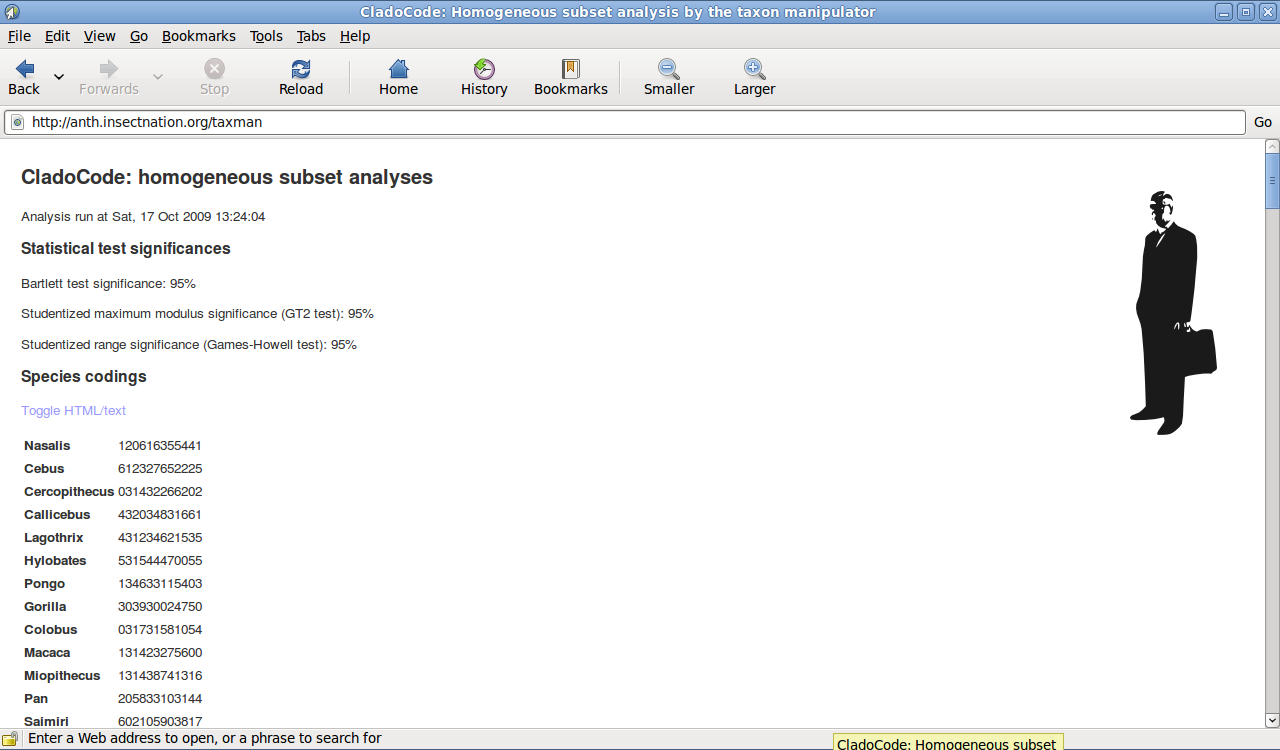}\\[2em]
  \includegraphics[width=0.95\textwidth]{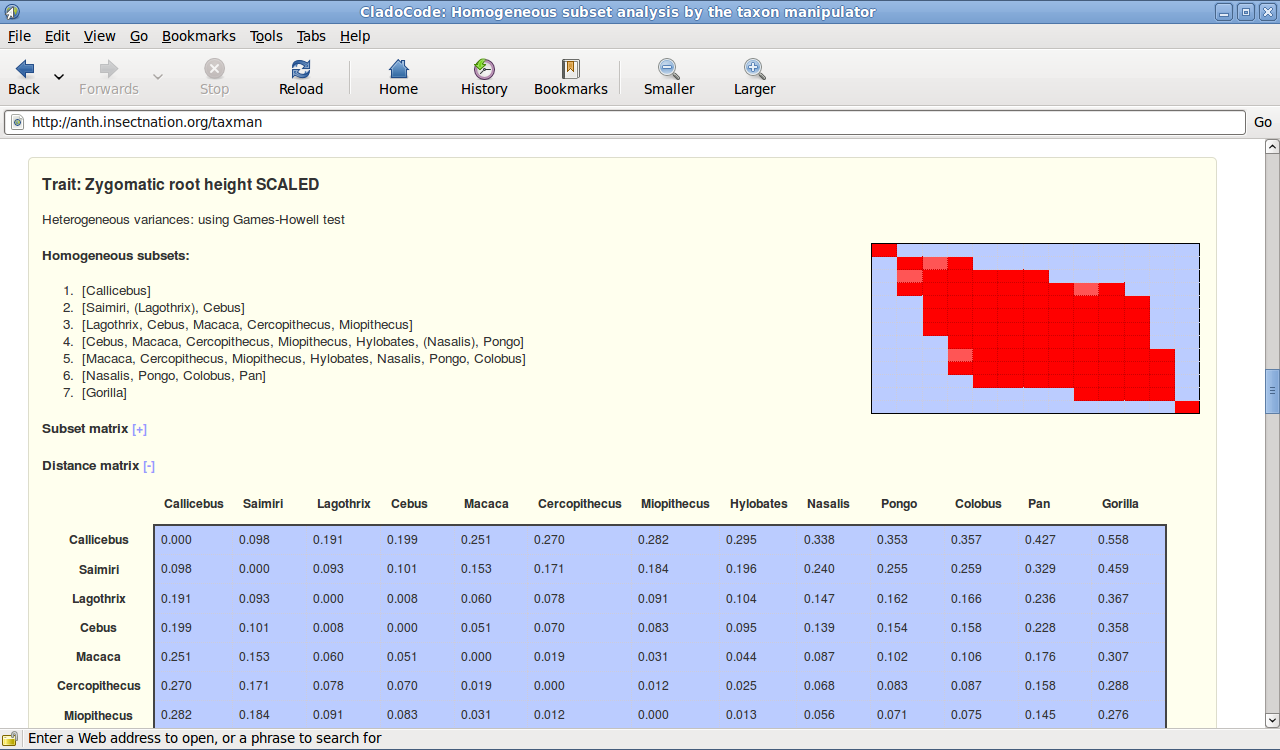}
  \caption{Screen shots of a) the TaxMan main screen, b) the main results screen
    from the analysis of data from anthropoid primates (see text) and c)
    specific results for one trait in the above data, nasoalveolar height. In
    shot a), tab-delimited plain text files are selected by clicking on the
    ``Browse'' button and highlighting the relevant local file; the file is then
    analysed by clicking on the ``Upload'' data file button. The output
    describes the tests utilised, provides a coded matrix (HTML or plain text),
    and gives a detailed breakdown of the implementation for each character.}
  \label{fig:screenshots}
\end{figure}

To address this shortcoming, the present contribution outlines a new Web-based
package that can be used to code numerical data for phylogenetic analysis.
TaxMan (\url{http://anth.insectnation.org}) is an implementation of the
Homogeneous Subset Coding (HSC) method (Simon,~1983) that can be run via any Web
browser (Figure \ref{fig:screenshots}).  Analogous to ClustalW
(Higgins~et~al.,~1994), the Web-based sequence alignment site
(\url{http://www.ebi.ac.uk/Tools/clustalw2/}), TaxMan allows the entry of data
from which the algorithm will produce a data matrix amenable to tree-searching
computer software such as PAUP* (Swofford,~2003) or TNT (Goloboff~et~al.,~2003;
Goloboff~et~al.,~2008).

\section{Methods}

TaxMan proceeds by evaluating the data using standard statistical methods
(Sokal~and~Rohlf,~1981).  For each trait, homogeneity of variances is determined
with Bartlett's test.  Traits with homogeneous variances are then subjected to
the parametric GT2 multiple comparisons test.  Characters demonstrating
non-homogeneous variances are evaluated using the non-parametric Games and
Howell procedure.  As the multiple comparisons procedures are post-hoc,
two-tailed tests are used throughout.  The standard critical value of $p < 0.05$
is utilised for all of the calculations.  Taxa are sorted into homogeneous
subsets (i.e., those taxa that do not differ significantly from each other), and
all taxa belonging to exactly the same subsets are assigned the same code.  The
procedure is identical to that of Simon~(1983), with the exception that codes
are represented by single alphanumeric characters only (\texttt{0-9},
\texttt{a-z}, and \texttt{A-Z}). The resulting data matrix is sorted with taxa
arranged vertically and characters positioned along the horizontal axis.
Details of the statistical algorithms and their implementation are given in the
Appendix.

\section{Worked example}

\begin{figure}
  \centering
  \includegraphics[width=0.7\textwidth]{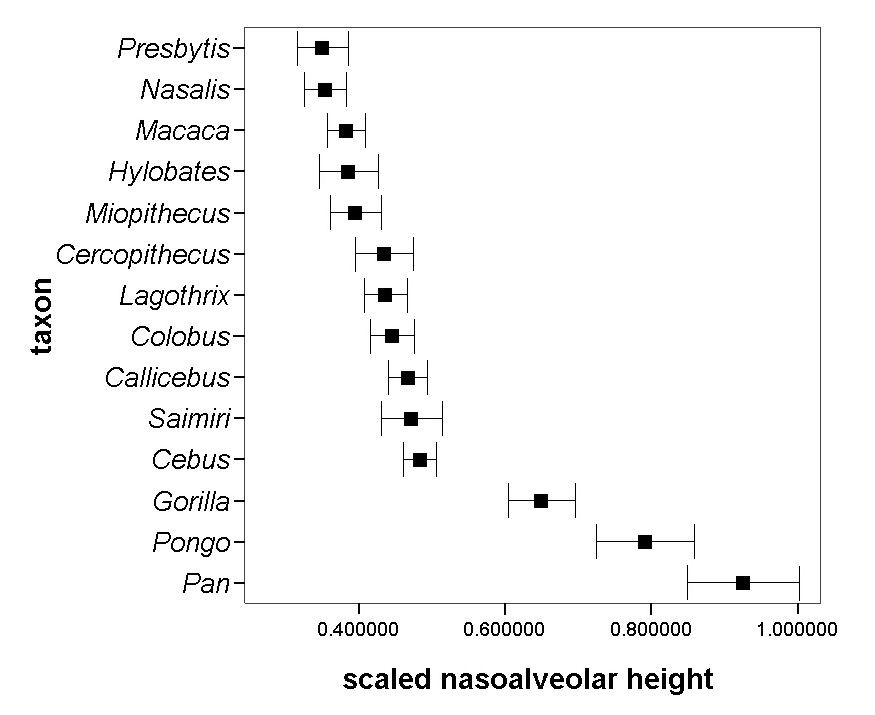}
  \caption{Distribution of nasoalveolar height in a sample of anthropoid
    primates scaled by grand mean.  For each taxon, the mean is represented by
    the filled square, and the horizontal line gives the 95\% confidence
    interval of the mean.}
  \label{fig:neheightdbn}
\end{figure}

To illustrate the applicability of TaxMan, a linear measurement of the facial
skeleton of anthropoids (Mammalia:~Primates), previously utilised in
phylogenetic analysis of the group (Rae,~1993), is analysed below.  The length
of the nasoalveolar clivus (measured from prosthion to the anterior nasal spine)
has been hypothesised to distinguish great apes (Hominidae, including the extant
genera \taxon{Pongo}, \taxon{Pan}, \taxon{Gorilla}, and \taxon{Homo}) from other
anthropoid primates (Harrison,~1982; Rae,~1997; Rae,~1999).  Data from 19--20
individuals each from 15 extant genera were taken with digital sliding
callipers accurate to \unit{0.01}{\milli\metre}.  The individual variates were
scaled by dividing each value by a grand mean of 12 measurements taken from the
same individual (Jungers~et~al.,~1995).  The scaled variates
(Figure~\ref{fig:neheightdbn}) were coded via TaxMan, and compared with codings
and treatments from both Morphocode, which implements a gap-weighting coding
algorithm (Thiele,~1993), and the continuous data option (Goloboff~et~al.,~2004)
of TNT.  Morphocode allows the number of character states in the coding to vary
between the two default settings (10, 26); only these two settings were used.
The ``cont'' algorithm of TNT (Goloboff~et~al.,~2003) does not code data as
such, so two alternative ways of entering continuous data allowed by the program
(means, ranges) were implemented.  These were added to an identical string of
identical dummy variables, so that the resulting cladograms (and character
weightings) differ only due to the influence of the single continuous
character. Descriptive statistics were derived using SPSS for Windows,
version~15.0 (SPSS~Inc.,~Chicago).

The coding of the scaled data by the various methods is given in Table
\ref{table:neheightdbn}. The results from TaxMan consist of a single set of
codes for the taxa analysed.  There are six states (0--5) and the hominids are
characterised by an elevated value.  The mean values were analysed with
Morphocode, using the $n=10$ states and $n=26$ states settings. As Morphocode
uses non-alphanumeric characters as codes between the integers and letters, to
facilitate the comparison with TaxMan all states greater than 9 were translated
into their alphanumeric equivalents.  It is obvious that the number of states
chosen affects the resulting states directly; whereas \taxon{Callicebus} and
\taxon{Cebus} were deemed to be the ``same'' in the $n=10$ states coding (both
state 2), they are ``different'' in $n=26$ states analysis (states 5 and 6,
respectively).  Similarly, \taxon{Cercopithecus}, \taxon{Colobus},
\taxon{Hylobates}, \taxon{Lagothrix}, \taxon{Macaca}, and
\taxon{Miopithecus} are indistinguishable in the first run and spread across
three states in the second.  Other alternative codings can be obtained by
choosing intermediate values for the number of states.  All of these
distinctions are arbitrary and have nothing to do with the morphometric
qualities of the organisms.

\begin{figure}
  \centering
  \includegraphics[width=0.45\textwidth]{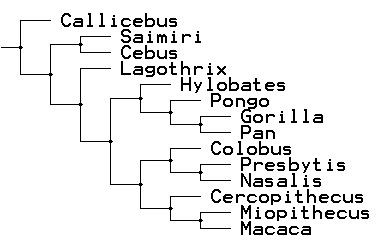}
  \includegraphics[width=0.45\textwidth]{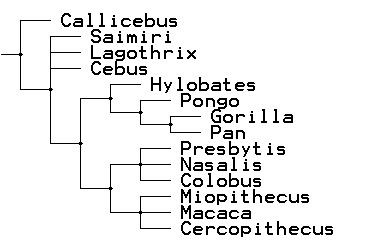}
  \caption{Topologies and character steps for nasoalveolar height data analysed
    as continuous data in TNT., using a) mean values only, and b) ranges of the
    mean one standard deviation. The topology shown in b) is identical to the
    most parsimonious arrangement of the matrix when analysed excluding the
    metric trait. The one continuous character contributes alternately 0.877
    steps in a) and 0.358 steps in b).  Thus, the (arbitrary) choice of how to
    characterise the continuous data has a discernable effect on the resulting
    weight of the trait across the topology and on the resolution of the
    resulting most parsimonious topology.}
  \label{fig:neheighttopos}
\end{figure}

Two additional tests were conducted by adding alternatively a) taxon means and
b) 68\% confidence intervals (mean $\pm$ one standard deviation) to an
artificial matrix (Table \ref{table:artificial}) for analysis in TNT using the
``nstates cont'' option (Figure \ref{fig:neheighttopos}).  This method does not
allow a direct comparison to TaxMan, as there are no codings per se to contrast,
but the two options used produce different topologies and numbers of steps
produced for the one continuous character.  The ``means'' run showed more
resolution, splitting several polytomies seen in the ``ranges'' run, implying
that the former strategy produces synapomorphies not discovered in the latter.
Also, the continuous character introduced more steps (or fractions thereof) in
the ``means'' run (0.877 steps) than the ``range'' run (0.358 steps).  Thus,
although the data had not changed, both the Morphocode and the ``analyzing
continuous characters as such'' methods produce results that differ, depending
on the arbitrary choice of number of states or method of entry.

\section{Discussion}

Much has been made, by workers advocating coding methods for phylogenetics, of
the observation that including measurement data in phylogenetic analyses
increases information content (Chappill,~1989).  While this is undoubtedly true,
it is probably trivial; the inclusion of any additional data could also increase
the resolution/support for various clades.  As was found with PTP tests
(Slowinski~and~Corhter,~1998) and other measures of ``phylogenetic content'',
there are likely to be very few observations of organisms that cannot provide
phylogenetic information of some sort.  The central issue of coding should be
that it provides a way of including metric data without resorting to arbitrary
thresholds or ranges.

One objection that has been raised with respect to coding methods is that they
may allow the separation of taxa with mean values for a particular character
that are not significantly different from one another (Farris,~1990).  In HSC,
this is implemented via the coding of intermediates: if the means of taxa A and
C are significantly different for a trait, but taxon B has a mean that lies
between those of A and C, but is not significantly different from either, taxon
B is coded as intermediate.  As pointed out previously (Swiderski~et~al.,~1998),
this is exactly the distribution that would be expected in a trait changing via
gradualism; to argue that this ``creates non-existent differences'' is to ignore
the obvious power of selection on populations.  The alternative in multiple
comparisons is to place all three in the same category, even though demonstrable
differences occur within.  The default assumption should be that any information
on distinctions between taxa is relevant for phylogeny and should be included,
rather than ignored.

A second but related issue in cladistic coding for involves the scaling of
metric traits.  An organism's size is one of its most fundamental adaptations
(Jungers,~1985).  A good deal of variance in any metric assessment of a group of
organisms will be directly attributable to size alone.  Raw (or even incorrectly
scaled) measurements will introduce an element of non-independence in cladistic
analysis (Rae,~2002), violating one of the fundamental assumptions of the
methods of phylogenetic systematics (Farris,~1983) and resulting in what amount
to single-trait phylogenies.  Spurious conclusions (Collard~and~Wood,~2000) can
result from failing to take into account the effect of size properly
(Gilbert~and~Rossie,~2007).  TaxMan assumes that scaling of the data has already
taken place, using a method consistent with character independence.

At the heart of the phylogenetic systematics is the assumption that the
characters utilised consist of heritable states that can be differentiated in an
accurate and replicable manner, such that, for the same data, the same
information is conveyed.  In addition, history teaches us that methods that are
easy to use will be, regardless of their appropriateness; a child with a hammer
views the whole world as a nail.  As the methods implemented in TaxMan perform
character coding in a theoretically justifiable, consistent, and non-arbitrary
way, it should provide the means for the inclusion of valuable data in cladistic
analysis in a way that is relatively simple, which in turn will increase our
understanding of the pattern of the tree of life.

\begin{table}
  \centering
  \caption{Alternative coding schemes for scaled nasoalveolar height in anthropoid primates.}
  \vspace*{5mm}
  \begin{tabular}[h]{lccccc}
    \toprule
    Taxon  & Mean & 68\% conf. interval & TaxMan & \multicolumn{2}{c}{Morphocode} \\
    \cmidrule{5-6}
           &    &    &    & $n=10$ & $n=26$ \\
    \midrule
    \taxon{Callicebus}    & 0.47 & 0.41--0.53 & 1 & 2 & 5 \\
    \taxon{Cebus}         & 0.48 & 0.44--0.53 & 2 & 2 & 6 \\
    \taxon{Cercopithecus} & 0.44 & 0.35--0.52 & 1 & 1 & 4 \\
    \taxon{Colobus}       & 0.45 & 0.38--0.51 & 1 & 1 & 4 \\
    \taxon{Gorilla}       & 0.65 & 0.56--0.75 & 3 & 5 & d \\
    \taxon{Hylobates}     & 0.39 & 0.30--0.47 & 1 & 1 & 2 \\
    \taxon{Lagothrix}     & 0.44 & 0.38--0.50 & 1 & 1 & 4 \\
    \taxon{Macaca}        & 0.38 & 0.33--0.44 & 1 & 1 & 1 \\
    \taxon{Miopithecus}   & 0.40 & 0.32--0.47 & 1 & 1 & 2 \\
    \taxon{Nasalis}       & 0.35 & 0.29--0.42 & 0 & 0 & 0 \\
    \taxon{Pan}           & 0.93 & 0.77--1.08 & 5 & 9 & p \\
    \taxon{Pongo}         & 0.79 & 0.66--0.93 & 4 & 7 & j \\
    \taxon{Presbytis}     & 0.35 & 0.28--0.43 & 0 & 0 & 0 \\
    \taxon{Saimiri}       & 0.47 & 0.38--0.56 & 1 & 2 & 5 \\
    \bottomrule
  \end{tabular}
  \label{table:neheightdbn}
\end{table}

\begin{table}
  \centering
  \caption{Artificial character matrix used to evaluate different data
    input methods for the ``nstates cont'' option in TNT.}
  \vspace*{5mm}
  \begin{tabular}[h]{lcccccccccccccc}
    \toprule
    Taxon  & &&&&&&&&&&&&& \\
    \midrule
    \taxon{Callicebus}     & 0 & 0 & 0 & 0 & 0 & 0 & 0 & 0 & 0 & 0 & 0 & 0 & 0 & 0 \\
    \taxon{Cebus}          & 0 & 0 & 0 & 0 & 0 & 0 & 0 & 0 & 0 & 0 & 0 & 0 & 0 & 0 \\
    \taxon{Cercopithecus}  & 1 & 1 & 1 & 1 & 1 & 1 & 0 & 0 & 0 & 0 & 0 & 0 & 0 & 0 \\
    \taxon{Colobus}        & 1 & 1 & 1 & 1 & 0 & 0 & 1 & 1 & 0 & 0 & 0 & 0 & 0 & 0 \\
    \taxon{Gorilla}        & 1 & 1 & 0 & 0 & 0 & 0 & 0 & 0 & 1 & 1 & 1 & 1 & 1 & 1 \\
    \taxon{Hylobates}      & 1 & 1 & 0 & 0 & 0 & 0 & 0 & 0 & 1 & 1 & 0 & 0 & 0 & 0 \\
    \taxon{Lagothrix}      & 0 & 0 & 0 & 0 & 0 & 0 & 0 & 0 & 0 & 0 & 0 & 0 & 0 & 0 \\
    \taxon{Macaca}         & 1 & 1 & 1 & 1 & 1 & 1 & 0 & 0 & 0 & 0 & 0 & 0 & 0 & 0 \\
    \taxon{Miopithecus}    & 1 & 1 & 1 & 1 & 1 & 1 & 0 & 0 & 0 & 0 & 0 & 0 & 0 & 0 \\
    \taxon{Nasalis}        & 1 & 1 & 1 & 1 & 0 & 0 & 1 & 1 & 0 & 0 & 0 & 0 & 0 & 0 \\
    \taxon{Pan}            & 1 & 1 & 0 & 0 & 0 & 0 & 0 & 0 & 1 & 1 & 1 & 1 & 1 & 1 \\
    \taxon{Pongo}          & 1 & 1 & 0 & 0 & 0 & 0 & 0 & 0 & 1 & 1 & 1 & 1 & 0 & 0 \\
    \taxon{Presbytis}      & 1 & 1 & 1 & 1 & 0 & 0 & 1 & 1 & 0 & 0 & 0 & 0 & 0 & 0 \\
    \taxon{Saimiri}        & 0 & 0 & 0 & 0 & 0 & 0 & 0 & 0 & 0 & 0 & 0 & 0 & 0 & 0 \\
    \bottomrule
  \end{tabular}
  \label{table:artificial}
\end{table}

\break

\appendix

\clearpage
\section{Classification algorithm}
\label{app:alg}

Encoding an informally specified procedure in strict algorithmic form is an
error-prone process, hence we present here a formal description of the TaxMan
classification algorithm. Since the TaxMan procedure treats each trait in
isolation, it is sufficient to describe the procedure for a single trait, with
the understanding that this will be repeated for each of the $T$ traits.

\newcommand{\numS}[1][s]{\ensuremath{N^{(#1)}}\xspace}
\newcommand{\dofS}[1][s]{\ensuremath{n^{(#1)}}\xspace}
\newcommand{\xS}[2][s]{\ensuremath{x^{(#1)}_{#2}}\xspace}
\newcommand{\meanS}[1][s]{\ensuremath{\langle x^{(#1)} \rangle}\xspace}
\newcommand{\sdS}[1][s]{\ensuremath{\sigma^{(#1)}}\xspace}
\newcommand{\varS}[1][s]{\ensuremath{\sigma^2_{(#1)}}\xspace}

\paragraph{1. Data entry}
For each taxon $s$ of $S$ taxa with measurements for trait $t$, record the
$\numS$ trait measurements, $\xS{i}$.

\paragraph{2. Compute Bartlett statistic}
Initially, we must characterise the set of datasets according to comparison of
the variances of trait data for the different taxa. To do this, we calculate
the Bartlett statistic, defined as
\begin{equation}
  \label{eq:bartlett}
  B =
  \frac{1}{f}
  \left[
  \left( \sum_s \dofS \right)
  \log\left( \frac{ \sum_s \dofS \varS }{ \sum_s \dofS } \right) -
  \sum_s \dofS \log\varS
  \right]
\end{equation}
where the degrees of freedom $\dofS = \numS-1$, and the correction factor
$f$ is defined as
\begin{equation}
  \label{eq:bartlettcorrfactor}
  f =
  1 +
  \frac{1}{3(S-1)}
  \left(
    \sum_{s} \frac{1}{\dofS} -
    \frac{1}{ \sum_{s} \dofS }
  \right)
\end{equation}

\newcommand{\conf}{\ensuremath{\mathcal{C}}}

\paragraph{3. Test for homogeneity of variances}
The Bartlett statistic is then compared to the $\chi^2$ distribution to
determine whether the datasets are to be considered to have homogeneous
variances. This is the case if $B < \chi^2(\conf_B, S-1)$.

\newcommand{\dists}[1][ij]{\ensuremath{d_{#1}}\xspace}
\newcommand{\critdists}[1][ij]{\ensuremath{D_{#1}}\xspace}
\newcommand{\booldists}[1][ij]{\ensuremath{\mathcal{D}_{#1}}\xspace}

\paragraph{4. Calculate distance matrix}
We now build a matrix of the statistical distances between each
taxon's mean trait value, i.e.
\begin{equation}
  \label{eq:dists}
  \dists =
  | \meanS[i] - \meanS[j] |
\end{equation}
for all taxa $i,j$ in $\{s\}$. Note that by construction this matrix is
symmetric ($\dists[ij] = \dists[ji]$) and the diagonal elements \booldists[ii]
are all zero.

\paragraph{5. Calculate critical distance matrix}
Given a distance matrix, we require a corresponding definition of a critical
distance above which to consider two taxa as being statistically separated (for
this particular trait). Such a critical distance must depend on the number of
samples for each dataset, and hence is itself a matrix, $D$, with elements
$D_{ij}$ corresponding to distances $d_{ij}$. The choice of critical distance
also depends on the result of the previously computed variance homogeneity:
datasets with homogeneous variances are treated as in 5(a), while inhomogeneous
variances require the treatment in 5(b).

\newcommand{\pvar}{\ensuremath{\langle \sigma^2 \rangle^\prime}\xspace}

\paragraph{5\,(a) Homogeneous case}
We compute the Hochberg GT2 critical distances, defined as
\begin{equation}
  \label{eq:gt2}
  \critdists^{\,\text{homo}} =
  \text{SMM}(k^\star, \textstyle\sum_s \dofS, \conf^{\text{SMM}})
  \sqrt{ \pvar \left( \frac{1}{\numS[i]} + \frac{1}{\numS[j]} \right) }
\end{equation}
where $\text{SMM}(k, d, \conf)$ is the Studentised maximum modulus distribution
for $k$ samples, $d$ degrees of freedom, and confidence level \conf; $k^\star
= S (S-1) / 2$; and the ``pooled variance'', $\pvar$, is defined as
\begin{equation}
  \label{eq:pooledvariance}
  \pvar = \frac{\sum_s \dofS \varS}{\sum_s \dofS}
\end{equation}

\newcommand{\meanvarS}[1][s]{\ensuremath{\hat{\sigma}^2_{(#1)}}\xspace}

\paragraph{5\,(b) Inhomogeneous case}
We compute the Games--Howell critical distances, expressed in terms of the
squared standard errors of the means, $\meanvarS = \varS / \numS$:
\begin{equation}
  \label{eq:gameshowell}
  \critdists^{\,\text{inhomo}} =
  \text{SR}(S, \nu^\star, \conf^{\text{SR}})
  \sqrt{\meanvarS[i] + \meanvarS[j]}
\end{equation}
where $\text{SR}(k, d, \conf)$ is the Studentised range distribution for
$k$ samples, $d$ degrees of freedom, and confidence level \conf; and
\begin{equation}
  \label{eq:gameshowelldof}
  \nu^\star = \frac%
  { \left( \meanvarS[i] + \meanvarS[j] \right)^2}
  { (\meanvarS[i])^2/\dofS[i] + (\meanvarS[j])^2/\dofS[j] }.
\end{equation}

It is useful to note that both treatments produce critical distance measures
$D_{ij}$ with terms which are qualitatively average standard errors on the means
of sample sets $i$ and $j$. This would drive the criteria for sample set
distinguishability to become more demanding (produce smaller $D$ values) as
\numS[i] and/or \numS[j] increase, were it not for the tempering influence of
the SMM and SR distributions. Accordingly the critical distance measures should
be sample-size invariant, provided the samples are large enough to be safe from
small sample fluctuations.

\paragraph{6. Calculate binary separation matrix}
We now build a matrix of binary values by comparing the distances \dists with
the critical distances \critdists, such that
\begin{equation}
  \label{eq:binarydists}
  \booldists =
  \begin{cases}
    1 & \text{if $\dists < \critdists$}\\
    0 & \text{otherwise}
  \end{cases}
\end{equation}

This matrix hence encodes a pair of taxa whose distributions for this trait are
statistically indistinguishable with a 1, and a pair whose traits are
significantly different with a 0. Of course, the particular symbols used are
unimportant.

Note that by the symmetry of the distance and critical distance measures under
exchange of $i$ and $j$ indices, both \dists and \critdists are symmetric
matrices and hence so is \booldists. As a consequence of the null diagonal
elements of \dists, the diagonal elements of the binary matrix are all 1,
i.e. $\booldists[ii] = 1$.

\paragraph{7. Identify trait subsets}
We are now at the point where we can attempt to identify trait subsets by
grouping blocks of the binary matrix \booldists together. First, we order the
indices of the matrix by increasing corresponding mean, such that $\meanS[1] \le
\meanS[2] \le \dotsb \le \meanS[S]$. The motivation for this is that homogeneous
subsets will usually have similar means and thus ordering in \meanS will
optimally separate the subsets, so that they form ``block-diagonal'' rectangles
in the \booldists matrix with as little overlap as possible between each
rectangular block.

\newcommand{\imax}{\ensuremath{i_{\text{max}}}\xspace}

The algorithm for extracting unique subsets from a matrix with this block
structure is relatively intricate, compared to what has come before. The general
aim is that for a row $i$ in \booldists, we expect that a subset will include
all taxa between $i$ and some $\imax \ge i$, and hence the first step is to
identify \imax. We do this by starting with a proposed $\imax' = S$, i.e. the
taxon with the largest trait mean, and then incrementing $\imax'$ back towards
$i$ until \booldists[i \imax] = 1. All elements between $i$ and \imax are then
added to the new homogeneous subset $H_i$.

A complication is that the block diagonal structure may be imperfect, since the
critical distances \critdists depend on the variances of $i$ and $j$. For
example, in a potential subset of 3 elements the outside members (1 and 3) could
have variances much larger than that of taxon 2. It is then possible for
$\booldists[13] = \booldists[31] = 1$, while the overlaps of either 1 or 3 with
2 are not significant, $\booldists[21] = \booldists[12] = 0$ and/or
$\booldists[32] = \booldists[23] = 0$. These ``gaps'' must be handled by the
subset identification algorithm, and the approach we have chosen to take is to
treat them as if they had been ``1''s, but to mark the corresponding taxon as an
``exceptional'' member of $H_i$. This is only justified on the basis that such
gaps are likely to be due to insufficient sample data, rather than a true
statistical deviation, and accordingly subsets with exceptional members should
be used to indicate where data should be re-checked, or more measurements taken.

\begin{figure}[t]
  \centering

\begin{tikzpicture}
  \usetikzlibrary{matrix}
  \usetikzlibrary{positioning}

  \matrix[matrix of nodes,left delimiter=(,right delimiter=)](boolmatrix)
  {
    1 & 0   & 0   & 0 & 0 \\
    0 & 1   & $*$ & 1 & 0 \\
    0 & $*$ & 1   & 1 & 0 \\
    0 & 1   & 1   & 1 & 1 \\
    0 & 0   & 0   & 1 & 1 \\
  };

  \node[right=of boolmatrix](toarrow){$\Rightarrow$};

  \matrix[matrix of nodes,left delimiter=(,right delimiter=),right=of toarrow](blockmatrix)
  {
    1 &     &     &   &   \\
      & 1   & $*$ & 1 &   \\
      & $*$ & 1   & 1 &   \\
      & 1   & 1   & 1 & 1 \\
      &     &     & 1 & 1 \\
  };

  \fill[gray!70,rounded corners=0.2mm] (blockmatrix-4-4.north west) rectangle (blockmatrix-5-5.south east);
  \draw[gray,rounded corners=0.2mm,dashed] (blockmatrix-5-5.north west) rectangle (blockmatrix-5-5.south east);
  \draw[gray,rounded corners=0.2mm] (blockmatrix-4-4.north west) rectangle (blockmatrix-5-5.south east);

  \fill[gray!60,rounded corners=0.2mm] (blockmatrix-2-2.north west) rectangle (blockmatrix-4-4.south east);
  \draw[gray,rounded corners=0.2mm,dashed] (blockmatrix-3-3.north west) rectangle (blockmatrix-4-4.south east);
  \draw[gray,rounded corners=0.2mm,dashed] (blockmatrix-4-4.north west) rectangle (blockmatrix-4-4.south east);
  \draw[gray,rounded corners=0.2mm] (blockmatrix-2-2.north west) rectangle (blockmatrix-4-4.south east);

  \fill[gray!50,rounded corners=0.2mm] (blockmatrix-1-1.north west) rectangle (blockmatrix-1-1.south east);
  \draw[gray,rounded corners=0.2mm] (blockmatrix-1-1.north west) rectangle (blockmatrix-1-1.south east);

  \draw[white] (blockmatrix-1-1.center) node {\textbf{1}};
  \draw[white] (blockmatrix-3-3.center) node {\textbf{2}};
  \draw[white] (blockmatrix-5-5.center) node {\textbf{4}};
  \draw[gray] (blockmatrix-2-3.center) node {$*$};
  \draw[gray] (blockmatrix-3-2.center) node {$*$};

  \node[right=of blockmatrix](tosubs){$\Rightarrow$};

  \node[right=0.4 of tosubs](subsets) {
    $\displaystyle
    \begin{aligned}
      H_1 &= \{1\} \\
      H_2 &= \{2,(3),4\} \\
      H_4 &= \{4,5\} 
    \end{aligned}$
  };

\end{tikzpicture}
  \caption{An example of how subset block matrix structure arises from a boolean
    distance matrix with exceptional ``gap'' elements. Note that the exceptional
    ``$*$'' elements are really ``0'' elements whose position is anomalous, so
    they are treated as ``1''s and corresponding taxon is marked in the final
    subset \#2. The dashed lines indicate the invalid degenerate subsets.}
  \label{fig:blockmatrix}
\end{figure}

Finally, we must ensure that the identified subsets are non-degenerate. By this
we mean that no subset can be entirely contained within another --- only the
maximal subsets are counted as distinct. This rule can be enforced either as the
subsets are being constructed, or by a post-hoc checking procedure.

The procedure described in this step is illustrated in Figure
\ref{fig:blockmatrix}, which includes both exceptional entries and removal of
degenerate subsets, and an implementation is discussed in section
\ref{app:subsetid}.

\paragraph{8. Trait-specific subset coding}
For each trait, a given taxon will lie in between one and $N_h$ homogeneous
subsets. The particular pattern of sets in which it lies can be used to define a
unique identifier code such that any two taxa with the same code lie in the same
subsets. The form of this code is then a matter of implementation --- inside a
programmatic implementation, retaining the full information about which subsets
a taxon lies in is feasible and useful, but for human comprehension a more
intuitive and compact representation is desirable.

\paragraph{9. Taxon subset coding}
In this step we finally combine the results of each trait-wise subset encoding
into a single code per taxon. This is a matter of simply concatenating each of
the $T$ trait-wise encodings together, and is again implementation dependent.

\section{Implementation details}

The procedure described above has been implemented as a program in the Python
programming language, called TaxMan (for \emph{tax}on \emph{man}ipulator), with
both Web and command line interfaces. The program is currently accessible via the Web at
\url{http://anth.insectnation.org}.

\subsection{Data input}
The user provides a plain text file to TaxMan for subset coding. The first
(``header'') line in the file is the set of $T$ trait names, separated by tab
characters (spaces are allowed in the names). All subsequent rows have $T+1$
columns, again tab-separated, with the first column containing the taxon name
and the remaining $T$ columns being the scaled trait values corresponding to
those in the first row. These data are then read into a nested Python dictionary
structure and held in memory during processing.

\subsection{Statistical distributions}
The $\chi^2$ distribution is obtained from the implementation in the Python
``SciPy'' library for all confidence levels. The SMM and SR distributions,
however, are not available in any existing code library, and cannot be computed
from a closed form algebraic expression. Accordingly, rather than attempt to
invert a complex expression with the corresponding numerical pitfalls, we use an
interpolation for these two distributions, based on tables from
Rohlf~and~Sokal,~1981.

In both cases, the available tables are for $\conf = 95\% \text{ and } 99\%$,
and indexed in sample size $k$ and degrees of freedom $d$. Since the
distributions are asymptotic for large $k$ and $d$, we use harmonic
interpolation in each direction, i.e. in one dimension for a function $f(x)$
with $x$ lying between anchor points $x_1$ and $x_2$,
\begin{equation}
  \label{eq:harmonicipol}
  f(x) \sim \left[ (1-a)\frac{1}{f(x_1)} + a\frac{1}{f(x_2)}  \right]^{-1}
\end{equation}
where $a = (x-x_1)/(x_2 - x_1)$. In two dimensions we need to compute extra
pseudo-anchor points. For $k_1 \ge k \ge k_2$ and $d_1 \ge d \ge d_2$, we can
first interpolate in $k$ twice to get the new anchors $f(k,d_1)$ and $f(k,d_2)$,
then interpolate between these in $d$ to obtain $f(k,d)$. Or we could swap the
order, so that $f(k_1,d)$ and $f(k_2,d)$ are the intermediate points. As it
happens, the order is unimportant, yielding
\begin{equation}
  \label{eq:2dharmonicipol}
  f(k,d) \sim
  \left[
    \frac{(1-a_k)(1-a_d)}{f(k_1,d_1)} +
    \frac{(1-a_k) \, a_d}{f(k_1,d_2)} +
    \frac{a_k(1-a_d)}{f(k_2,d_1)} +
    \frac{a_k \, a_d}{f(k_2,d_2)} +
  \right]^{-1}
\end{equation}
where $a_k$ and $a_d$ are defined as for $a$ in equation
\eqref{eq:harmonicipol}, with $x$ replaced by $k$ or $d$ as appropriate.

The final implementation detail is the treatment of boundaries. The tables
available to us contained $k \in [3,20]$ and $d \in [5, \infty]$ for the SMM,
and $k \in [2, 100]$, $d \in [1, \infty]$ for SR. As the distributions diverge
for low $k$ and $d$ values, we do not attempt to ``plug the gaps'' at the low
ends, but the asymptotic behaviour at high $k$ means that it is acceptable to
lock the value at that of the last anchor points. Hence,
$\text{SR}(k=200,d=4.5)$ ($d^{\,\text{SR}} = \nu^\star$ is not usually an integer) will actually
be treated as $\text{SR}(k=100,d=4.5)$, with negligible error.

\begin{figure}[t]
  \centering
  \subfigure[]{\label{fig:smm95}\includegraphics[width=0.45\textwidth]{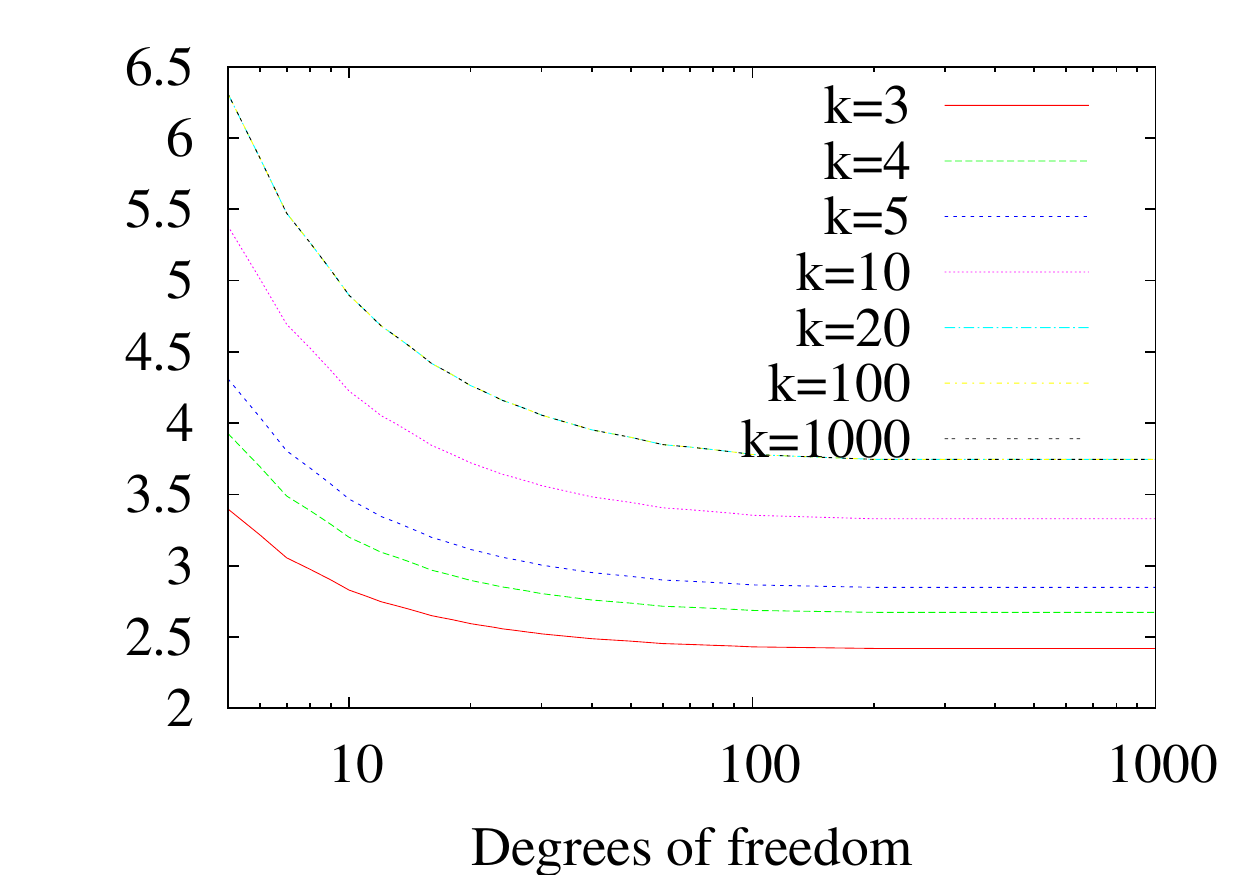}}
  \subfigure[]{\label{fig:smm99}\includegraphics[width=0.45\textwidth]{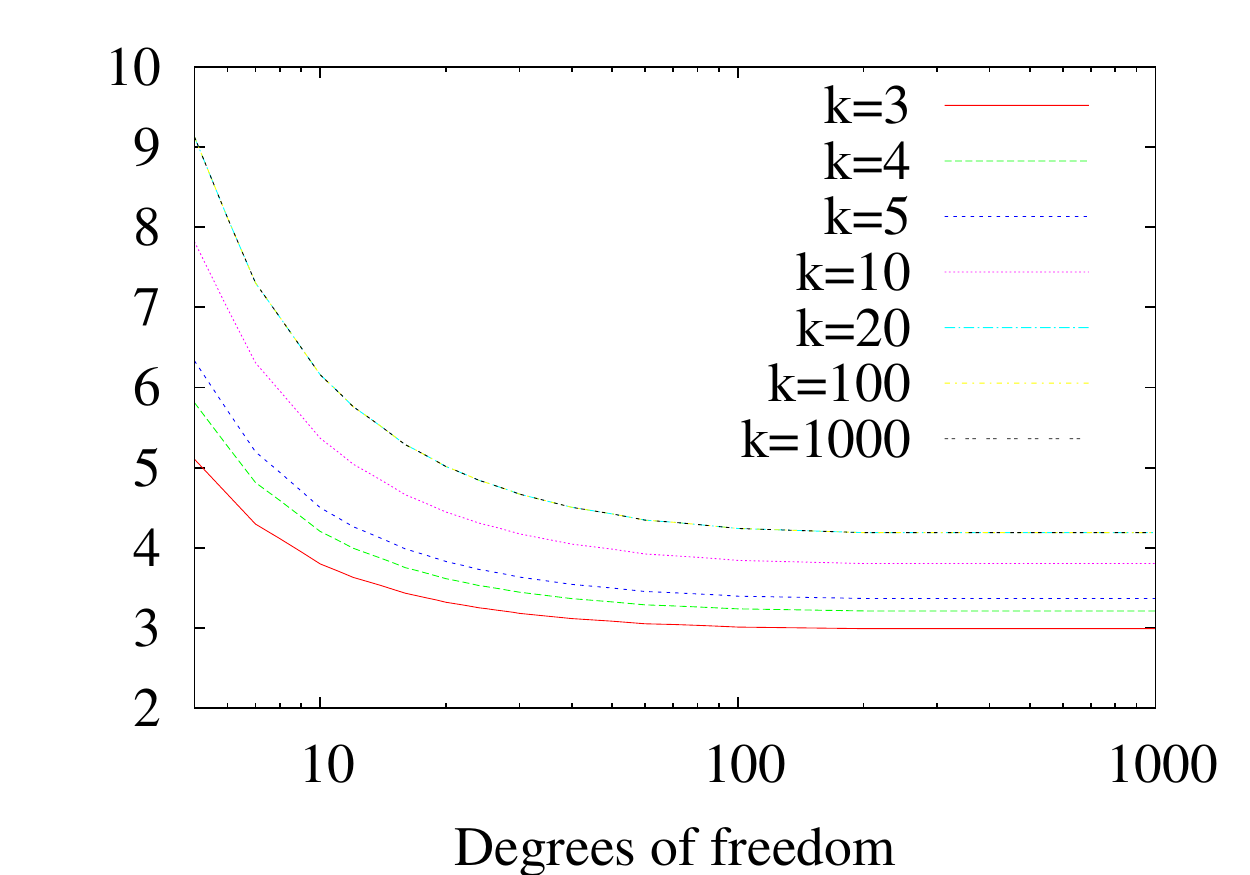}}
  \caption{Studentised maximum modulus distributions for confidence levels of
    95\% and 99\% respectively}
  \label{fig:smm}
\end{figure}

\begin{figure}[t]
  \centering
  \subfigure[]{\label{fig:sr95}\includegraphics[width=0.45\textwidth]{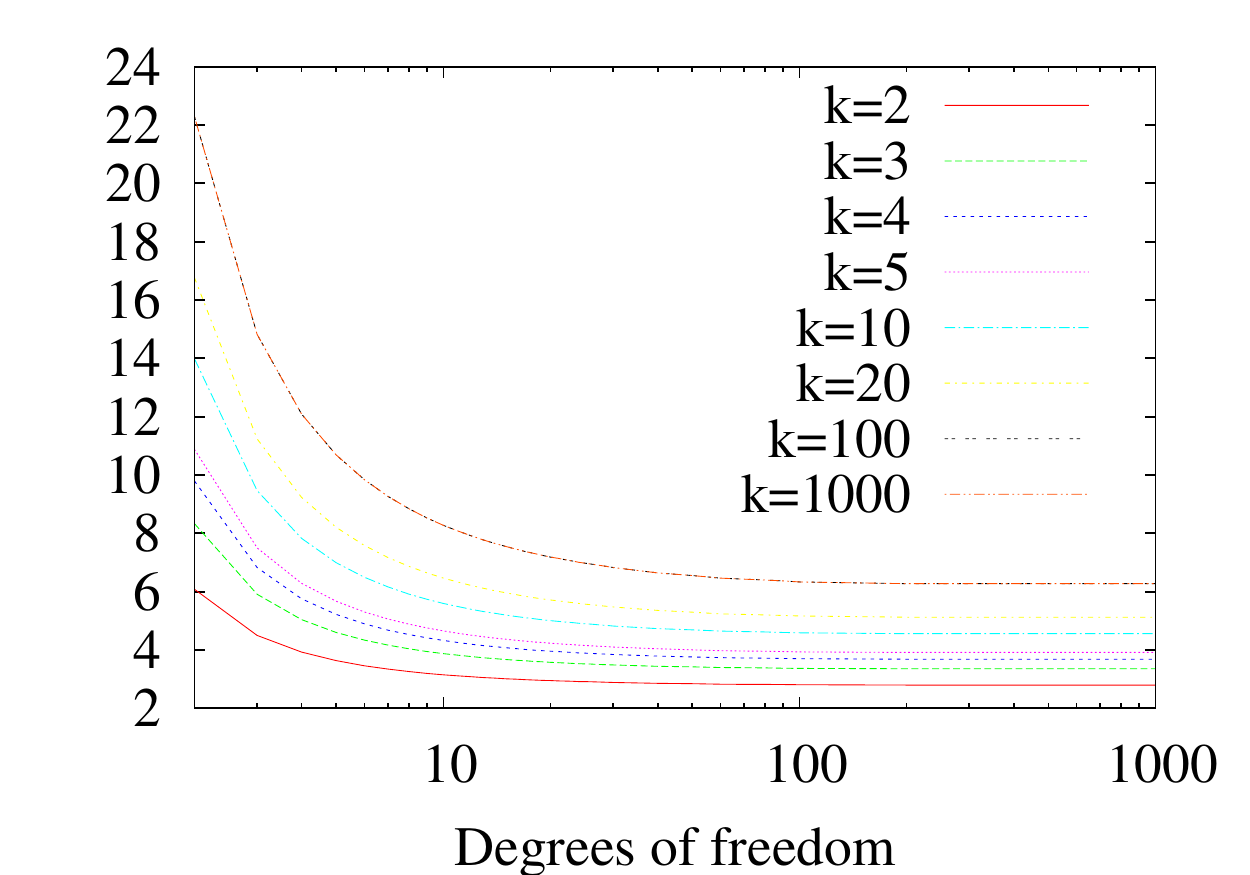}}
  \subfigure[]{\label{fig:sr99}\includegraphics[width=0.45\textwidth]{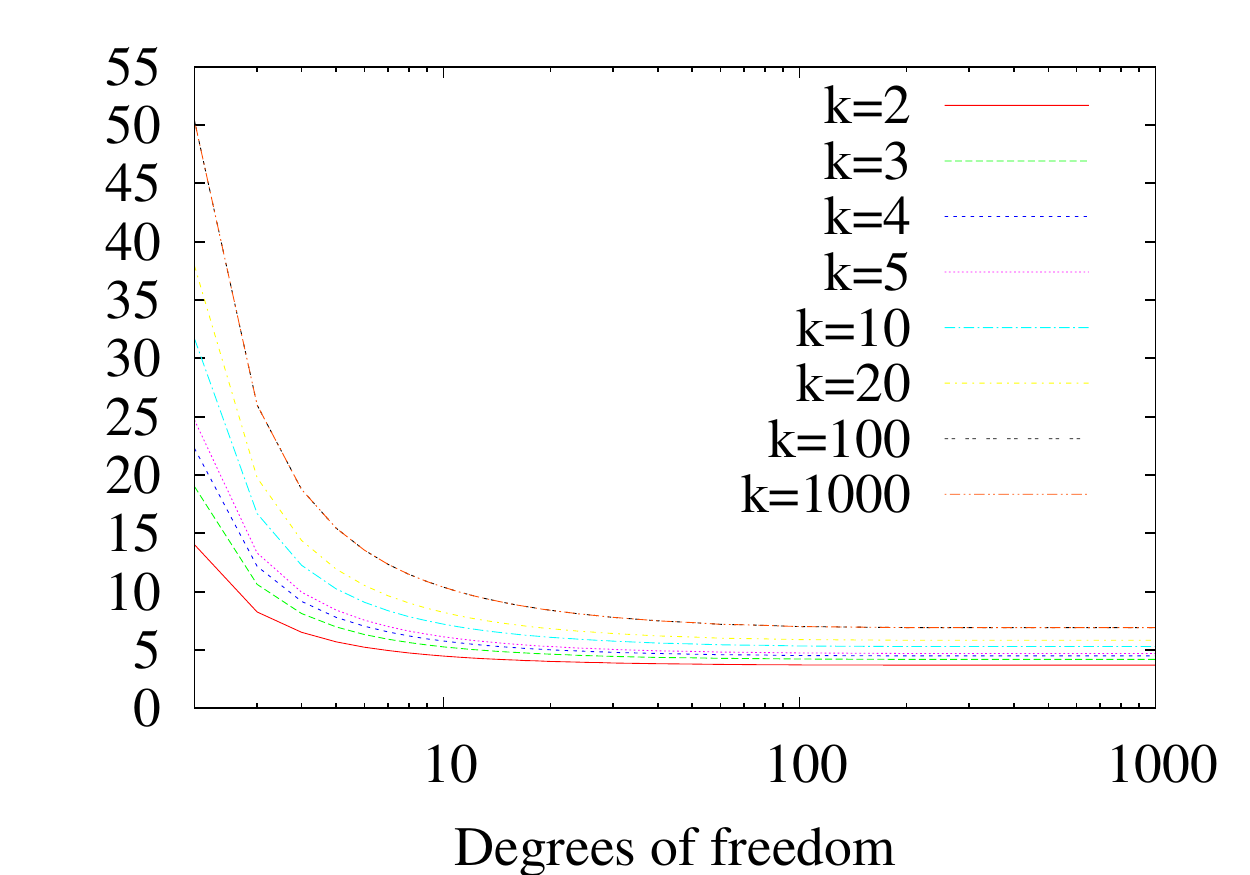}}
  \caption{Studentised range distributions for confidence levels of 95\% and
    99\% respectively}
  \label{fig:sr}
\end{figure}

As these distributions are not to our knowledge coded in any other programs or
code libraries, we have provided small programs in the TaxMan distribution which
can be used to obtain harmonic interpolations automatically. These may be useful
outside the context of TaxMan. The distributions themselves are plotted against
$d$ for various values of $k$ in Figures \ref{fig:smm} and \ref{fig:sr}.

\subsection{Subset identification}
\label{app:subsetid}
Identification of the subsets from the binary distance matrix, including
handling of exceptional set members, is performed as described in paragraph 7 of
Appendix \ref{app:alg}. However, since the procedure is relatively complex, we
include here a specific implementation in pseudo-code:

\vspace{7mm}
\goodbreak
{\smaller\sf\noindent
highestindex $:= S$\\
\textbf{for} $i$ \textbf{in} $\{1 \ldots S\}$:\\
\hphantom{xxx} $H_i := [\ ]$\\
\hphantom{xxx} foundmatch := False\\
\hphantom{xxx} highindex := 0\\
\hphantom{xxx} \emph{Count backwards until a match is found, then look for exceptions}\\
\hphantom{xxx} \textbf{for} $j$ \textbf{in} $\{S \ldots i\}$:\\
\hphantom{xxxxxx} \textbf{if} foundmatch = False \textbf{and} $\booldists[ij] = 1$:\\
\hphantom{xxxxxxxxx} highindex := $j$\\
\hphantom{xxxxxxxxx} foundmatch := True\\
\hphantom{xxxxxx} \textbf{if} foundmatch:\\
\hphantom{xxxxxxxxx} \textbf{add} $j$ \textbf{to} $H_i$\\
\hphantom{xxxxxx} \textbf{if} $\booldists[ij] = 0$ \textbf{and} (foundmatch \textbf{or} $j \le$  highestindex):\\
\hphantom{xxxxxxxxx} mark $j$ as an exceptional $H_i$ member\\
\hphantom{xxxxxxxxx} $\booldists[ij] := \booldists[ji] := *$\\
\\
\hphantom{xxx} \emph{Removing degenerate subsets}\\
\hphantom{xxx} \textbf{if} highindex $\le$ highestindex:\\
\hphantom{xxxxxx} \textbf{discard} $H_i$\\
\hphantom{xxx} \textbf{else}:\\
\hphantom{xxxxxx} highestindex := highindex
}

\subsection{Subset and species coding scheme}
We now consider the implementation of a concrete coding scheme. As described,
there are two stages: the assignment of per trait subset codes and the
combination of these into an overall species code.

The trait code is a case where it is tempting to think that a binary code will
work well, since a binary digit can be used per subset, which is ``0'' if a
taxon is not in the subset, and ``1'' if it is. However, this quickly leads to
large codes, since 10 subsets gives a code with 10 binary symbols: concatenating
such codes is unwieldy, even with conversion to base 10. A sequential numbering
scheme is more efficient, allowing single-character representations of subsets
in typical datasets. Sequential code numbers can be assigned using a simple
algorithm, such as the following:

\vspace{7mm}
\goodbreak
{\smaller\sf\noindent
highestindex $:= S$\\
\textbf{for} $s$ \textbf{in} $\{1 \ldots S\}$: \\
\hphantom{xxx} $H_s := [\ ]$; $C_s := [\ ]$\\
\textbf{for} $t$ \textbf{in} $\{ 1 \ldots T \}$:\\
\hphantom{xxx} code = -1\\
\hphantom{xxx} membership $:= [\ ]$\\
\hphantom{xxx} \textbf{for} $s$ \textbf{in} $\{ 1 \ldots S \}$:\\
\hphantom{xxxxxx} prevmembership := membership\\
\hphantom{xxxxxx} membership $:= [\ ]$\\
\hphantom{xxxxxx} \textbf{for} $h$ \textbf{in} $\{ 1 \ldots N_{\text{sets}} \}$:\\
\hphantom{xxxxxxxxx} \textbf{if} $s$ \textbf{is in} $H_h$:\\
\hphantom{xxxxxxxxxxxx} \textbf{add} setnum \textbf{to} membership\\
\hphantom{xxxxxx} \textbf{if} membership \textbf{is not} prevmembership:\\
\hphantom{xxxxxxxxx} code := code + 1\\
\hphantom{xxxxxx} \textbf{append} code \textbf{to} $C_s$\\
}

Having generated an ordered list of code numbers for each taxon, we convert this
list to a one-character-per-trait symbolic representation. We use an extension
of decimal digits to alphabetic characters, as for hexadecimal, such that every
code number maps to a single character in the 62-element list
$[0,1,2,\dots,9,\text{a},\text{b},\dots,\text{A},\text{B},\dots]$, so that,
e.g. character \#18 = ``h''. Each taxon's code list is thus converted into a
character string of length $T$, which is presented to the user at the end of
processing. Any user who wishes to program an extension to TaxMan will have the
full code objects available, rather than the string representation since the
latter is fundamentally just a presentation detail.





\end{document}